# Is Information in the Brain Represented in Continuous or Discrete Form?

James Tee, *Member, IEEE*, and Desmond P. Taylor, *Life Fellow, IEEE*

***Abstract*—The question of continuous-versus-discrete information representation in the brain is a fundamental yet unresolved question. Historically, most analyses assume a continuous representation without considering the discrete alternative. Our work explores the plausibility of both, answering the question from a communications systems engineering perspective. Using Shannon's communications theory, we posit that information in the brain is represented in discrete form. We address this hypothesis using 2 approaches. First, we identify the fundamental communication requirements of the brain. Second, we estimate the symbol error probability and channel capacity for a continuous information representation. Our work concludes that information cannot be communicated and represented reliably in the brain using a continuous representation – it has to be in a discrete form. This is a major demarcation from conventional and current wisdom. We apply this discrete result to the 4 major neural coding hypotheses, and illustrate the use of discrete ISI neural coding in analyzing electrophysiology experimental data. We further posit and illustrate a plausible direct link between Weber's Law and discrete neural coding. We end by outlining a number of key research questions on discrete neural coding.**

***Index Terms*—brain, coding, communications, computations, continuous, discrete, information, quantization, representation.**

## I. Introduction

Music consists of continuous vibrations of sound waves. These can be recorded (or stored) in either continuous (e.g., cassette tape, LP record) or discrete (e.g., CD, MP3) form. This leads to a fundamental research question in neuroscience, namely, is information in the brain represented in continuous or discrete form? To clarify, there is no doubt that neural signals (electrical waveforms) are continuous-valued (e.g., a typical cell membrane voltage potential can take any value between +40 mV and -70 mV); the question is then whether the information embedded within (or carried by) these neural signals is continuous or discrete. A helpful analogy is the Morse Code, where the information is clearly discrete (i.e., dots or dashes) but the electrical (or radio) signals that carry this discrete information are continuous. While this neural information representation question may seem trivial, it remains unresolved. McCulloch and Pitts [3] theorized that computation in the brain is digital (i.e., discrete), whereas Lashley [4] suggested that "*the brain is an analogical machine, not digital*". More recently, VanRullen and Koch [5] noted that a discrete representation of conscious perception (in the brain) is often "*considered but never widely accepted*" or indeed adopted. At the same time, a continuous representation "*cannot satisfactorily account for a large body of psychophysical data*". In the visual working memory literature, this continuous-versus-discrete debate is currently at an impasse – an extensive review of arguments for and against both positions can be found in Luck and Vogel [6], and Ma *et al.* [7]. Perhaps the problem is best summed up by Gallistel in 2016 [8]: "*We do not yet know in what abstract form (analog or digital) the mind stores the basic numerical quantities that give substance to the foundational abstractions, the information acquired from evidence.*" Note that the use of the term representation encompasses a broad range of information processing activities (or functions) in the brain, including thinking (performing computations, decision-making) and remembering (forming, storing and recollecting memories). In this paper, the terms continuous and analog are treated as equivalent in an engineering sense, as are the terms discrete and digital.

In the absence of a clear resolution to the debate, a continuous representation (of information in the brain) is almost always the default position when considering models and theories arising from experimental data. This is, in part, due to the appeal of the mathematical elegance and analytical convenience of continuous representations (or Real numbers). For example, Abbott, DePasquale and Memmesheimer [9] point out that neurons communicate with one another "*almost exclusively through discrete action potentials*" – yet, their models described in the rest of their paper employ continuous representation (i.e., derivative calculus using Real numbers). Similarly, Chaudhuri and Fiete [10] laud discrete attractor models for their superior stability and noise tolerance – yet, they default to continuous attractor models as the null hypothesis. A question to pose at this point is: why is this continuous-versus-discrete problem important to brain research?

One answer [5] is that a continuous representation "*cannot*

James Tee was previously with the Department of Psychology, New York University. He is now with the Communications Research Group, Department of Electrical & Computer Engineering, University of Canterbury, Private Bag 4800, Christchurch 8020, New Zealand (email: james.tee@canterbury.ac.nz).

Desmond P. Taylor is with the Communications Research Group, Department of Electrical & Computer Engineering, University of Canterbury, Private Bag 4800, Christchurch 8020, New Zealand (email: desmond.taylor@canterbury.ac.nz).

*satisfactorily account for a large body of psychophysical data*". For example, 1 cent does not have much value to most people. However, a person may decide to buy a product if priced at $1.99 – yet, refuse to buy the same product if priced 1 cent higher at $2.00. Such an abrupt (or step) change in the brain's purchasing decision cannot be modeled using a continuous representation despite extensive attempts to do so [11]. In contrast, models based on discrete representations of information are able to accommodate such abrupt (or step) changes. In the decision-making literature, Varshney and Varshney [12] showed that a Bayesian model using quantized (discrete) priors was better able to model racial discrimination behavior. In monetary economics, Khaw *et al.* [13] found that a discrete model was a better fit to data from a price-setting experiment than the so-called "optimal" (continuous) models. In mathematical psychology, Sun *et al.* [14] applied a Bayes-optimal quantized (discrete) model of perception to datasets of animal vocalizations and human speech, that closely mimicked the Weber-Fechner law. In theoretical neuroscience, Varshney *et al.* [15] posited that neurons with discrete-valued synaptic states might perform better than neurons with continuous-valued states. In a recent electrophysiology experiment, Latimer *et al.* [16] demonstrated that a discrete (stepping) model is a better fit to neural (macaque lateral intraparietal cortex) data than a continuous (ramping, diffusion-to-bound) model [17]. In machine learning, discrete (quantized or binary) neural networks have been shown to be superior to continuous neural networks in terms of error rate performance [18] and computation speed [19]. A further reason for favoring discrete representation, as suggested by Chaudhuri and Fiete [10], is noise tolerance. There are several sources of noise in the brain (e.g., sensory noise, cellular noise, motor noise) – an extensive review of these is found in Faisal *et al.* [20]. Due to the presence of noise, information that is represented or transported in continuous form inevitably degrades and becomes corrupted in an irrecoverable manner. Analogously, the quality of music that is recorded (or stored) in continuous (or analog) form (a cassette tape) will slightly degrade due to noise every time it is replayed. On the other hand, the quality of music that is recorded (or stored) in discrete form (CD) is robustly preserved with virtually no degradation on replay in the presence of noise.

In this paper, we address the discrete-versus-continuous question following 2 approaches, each drawn from communications systems engineering: 1) we identify the fundamental communication requirements of the brain; 2) we derive symbol error probability and channel capacity estimates for a continuous representation. In each instance, we consider the performance and plausibility of continuous versus discrete representation. Our findings demonstrate the superior performance of a discrete representation. We next apply the discrete representation results to 4 major hypotheses on neural coding and demonstrate the implications of these approaches. In addition, we illustrate how a discrete ISI neural coding scheme can be used to analyze and interpret electrophysiology experimental data. Finally, we extend the discrete representation results to Weber's Law. Overall, our results conclude that the brain must almost surely represent (and process) information in discrete form.

## II. Fundamental Communication Requirements of the Brain

We first outline 3 fundamental communication requirements of the brain, and their consequences.

1) <u>*Information in the brain must be transmitted reliably*</u>
Much information in the human sensory system must be transmitted over a relatively long distance while retaining its fidelity. One such example is vision. Information in human vision originates at the retina, is conveyed via the optic nerve to the Lateral Geniculate Nucleus (LGN) and arrives at the visual cortex, located at the back of the brain – a distance spanning almost the entire length of the brain [21]. The fidelity of this information must be retained throughout the transmission process.

2) <u>*Information in the brain must be stored and retrieved repeatedly*</u>
One example of this is memory. Our mind needs to be able to remember/recall/retrieve the same information (e.g., date of birth) repeatedly, possibly throughout an entire lifetime. Based on this recollection, we might decide on a desired course of action (e.g., how should I celebrate my birthday this year?). Without memory, there is no means to recall significant information and experiences in our lives. Thus, storage and repeated retrieval of information is crucial to the brain's function.

3) <u>*Noise exists and is ubiquitous in the brain*</u>
All electrical processes, including neural spikes, are noisy due to thermal fluctuations, known as thermal noise or Johnson noise [22]. There are also several additional sources of noise in the brain, such as sensory noise and motor noise [20]. Noise affects both information transmission and the retrieval of stored information (accessing memory) in 2 ways. First, noise corrupts information during transmission. Second, stored information (memory) can be compromised by the act of retrieving/accessing it. That is, accessing memory content risks altering it. Thus, one way or another, the effects of noise in brain communications must be mitigated.

A major consequence of these requirements is that information representation in the brain, in essence, cannot be continuous. If it were continuous (Real numbers which have infinite precision, where each value represents a unique message), then noise inevitably causes confusion of one message with another ($\text{message}_1$ + noise = $\text{message}_2$), resulting in information corruption no matter how large or small its magnitude (variance). We also note that, from a computer engineering standpoint, true Real number computers (i.e., employing continuous representation) are not physically realizable in principle, because there is an uncountable infinity of Real numbers. For example, the memory content of a modern digital computer can be concatenated to form a single long but finite-length word (e.g., an approximation to π); whereas true Real numbers (e.g., the true π) have infinite precision (decimal places), and consequently, require infinite space to store even a single Real number (or variable) which violates the basic foundation of information storage. Therefore, from a computer engineering standpoint, a truly continuous computational device is not physically realizable.

## III. Symbol Error Probability and Channel Capacity Estimates for Continuous Representation

All forms of communications employ a modulation scheme, that maps the source information onto a modulation scheme's signal constellation. The modulated signal is then transmitted over a noisy channel and demodulated at the receiver (Figure 1). Furthermore, a discrete information source must employ a digital (or discrete) modulation scheme – and similarly, a continuous information source must employ a continuous modulation scheme. Note that our use of the term signal arises from the perspective of Shannon communications theory, in the sense that a (or the) key purpose (or function) of a (modulation) signal is to carry, convey and/or represent the information that is to be communicated from the transmitter to the receiver. That is, a signal is a physical, quantifiable and measurable quantity. By its nature of physical existence, a signal is (almost) always continuous (or analog) in representation. Strictly speaking, in a physical system, there are no discrete signals per se, because all physical signals (e.g., electrical signals, optical signals) are continuous in nature. The signal, however, is distinct in definition from the information. Information is the message that is being carried, conveyed and/or represented by the signal, but information is not the signal itself.

Following communications theory, we know that the symbol error probability and channel capacity can be estimated for any digital (i.e., discrete) modulation scheme, across any noisy channel that has a defined statistical distribution. These communications principles are applicable to the brain, since all neural communication is electric in a very broad sense (e.g., neural spikes). With the aim of resolving the continuous-versus-discrete information representation debate, we approach the problem by estimating the symbol error probability and channel capacity for a continuous modulation scheme. The main idea is that, since a continuous information source requires a continuous modulation scheme, estimates of the symbol error probability and channel capacity of a continuous modulation scheme will provide a crucial and requisite indication concerning the plausibility and feasibility of a continuous modulation scheme, which in turn reflects on the plausibility and feasibility of a continuous information source. We note here that Shannon's seminal work on the source coding and channel coding theorems [23] focused solely on discrete information sources. To the best of our knowledge, the symbol error probability and channel capacity for a continuous information source have not been exactly estimated to date. This is in part due to the lack of a need to focus on a continuous information source since almost all modern information sources are discrete.

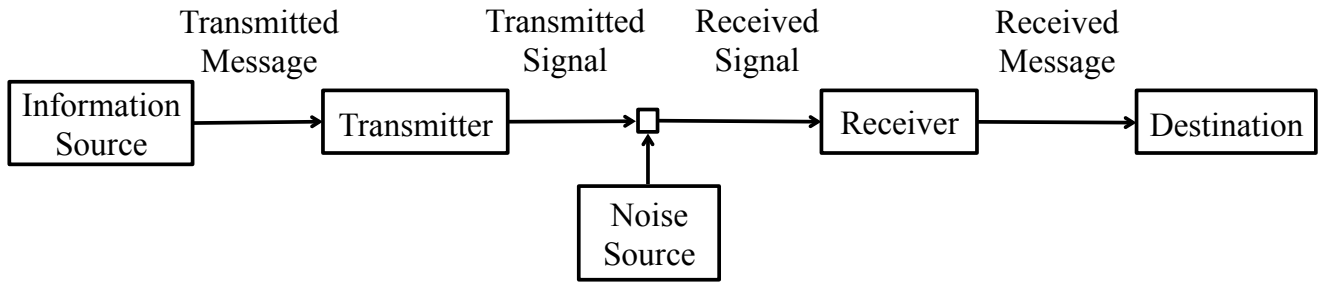

***Figure 1:*** *Classical model of a communications system* [23].

Using the symbol error probability estimate for a digital modulation scheme as a starting point, we increase the size of the signal constellation to the limit of infinity in order to develop an asymptotic estimate of the symbol error probability for a continuous modulation scheme. For simplicity, we employ, as a starting point, the simplest form of digital linear modulation – a one-dimensional Pulse Amplitude Modulation with $M$ levels ($M$-PAM) scheme (see Figure 2).

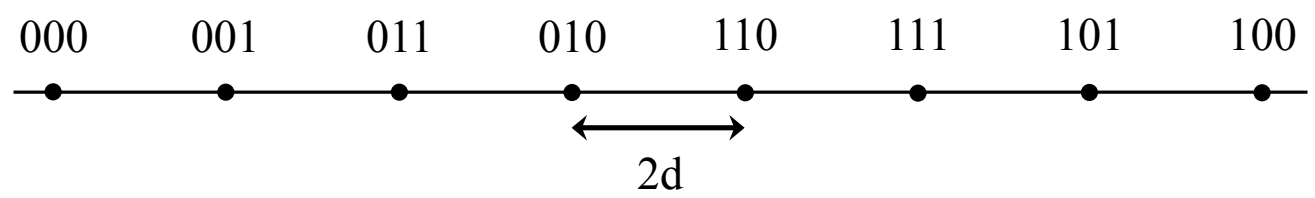

***Figure 2:*** *Gray encoded 8-PAM signal constellation.*

A pertinent question worth posing at this point is whether or not amplitude modulation can be realistically applied to analyzing the brain. That is, is there any evidence that the brain might be employing some form of amplitude modulation? The answer to this question is yes. Sensory and chemical synaptic input signals to neurons employ continuous amplitude representations, also known as graded potentials or graded voltage signals [24]. Such non-spiking neurons use graded potentials (i.e., continuous amplitude signals) for information transmission [25] [26]. For example, neurons in the outer retina process light stimuli using graded membrane potentials [27], as do neurons in the first stages of hearing [28]. Information at a visual synapse is also thought to be transmitted using an amplitude-modulated code [28].

We adopt the definitions used by Goldsmith [29], where $2d$ is the minimum distance between adjacent modulation symbols, $M$ is the number of possible transmit amplitudes (i.e., size of modulation signal set), and $\log_2 M$ is the number of bits per symbol. Based on these definitions, the symbol error probability for $M$-PAM in an Additive White Gaussian Noise (AWGN) channel is [29]:

$$P_S = \frac{2(M-1)}{M} Q\left(\sqrt{\frac{2d^2}{N_0}}\right)$$

where $Q(.)$ is the Gaussian Q-function and $N_0$ is the noise power spectral density. Note that, $Q(x)$ is the probability that a standard Gaussian random variable takes a value larger than $x$, and it is also related to the cumulative distribution function of the standard Gaussian distribution [29]. In the case of a continuous modulation scheme, $M$ is infinite and $2d$ is zero. Taking the symbol error probability to the limit as $M$ approaches infinity and $2d$ approaches zero (i.e., an approximation to the case of continuous modulation), we get:

$$\begin{aligned} \underset{\substack{\lim M\to\infty \\ d\to 0}}{P_S} &= \frac{2(\infty-1)}{\infty} Q\left(\sqrt{\frac{2(0)^2}{N_0}}\right) \\ &= 2(1)Q(0) \\ &= 2(0.5) \\ &= 1 \end{aligned}$$

This simple result shows that, in the presence of noise (AWGN), transmitting continuous information using a continuous $M$-PAM scheme results in errors in the entire transmitted message, with 100% certainty. In other words, no information will be successfully transmitted.

The average energy per symbol for $M$-PAM is [29]:

$$\overline{E_S} = \frac{(M^2-1)d^2}{3}$$

The average signal-to-noise ratio (SNR) per symbol is:

$$\overline{\gamma_S} = \frac{\overline{E_S}}{N_0} = \frac{(M^2-1)d^2}{3N_0}$$

Shannon's channel capacity for $M$-PAM in AWGN [23] is given by:

$$\begin{aligned} C &= B\log_2(1+\overline{\gamma_S}) \\ &= B\log_2\left(1+\frac{(M^2-1)d^2}{3N_0}\right) \end{aligned}$$

where $B$ is the bandwidth of the channel in Hz and $C$ is the channel capacity in bits/second. As $M$ approaches infinity and $2d$ approaches zero (i.e., an approximation to the case of continuous modulation), we get:

$$\begin{aligned} \underset{\substack{\lim M\to\infty \\ d\to 0}}{C} &= B\log_2\left(1+\frac{(M^2-1)d^2}{3N_0}\right) \\ &= B\log_2\left(1+\frac{\infty(0)^2}{3N_0}\right) \\ &= B\log_2(1) \\ &= 0 \end{aligned}$$

This implies that channel capacity approaches zero for the case of continuous PAM. By extension, then the channel capacity is actually zero for the case of a continuous information source. Therefore, a continuous representation of information in the brain is neither plausible nor feasible.

It may be obvious to a reader that the case of $M$-PAM with $M$ approaching infinity is essentially the same as AM radio, which is still in common usage today. A relevant question is: how could AM radio be operationally useful if the symbol error probability is one and channel capacity is zero? Or, did we make a fundamental mistake in our mathematical derivations? The answer, in our view, is that analog broadcast radio stations such as AM radio are rare exceptions. This is because the information they carry is analog (i.e., continuous) sound waves (i.e., audio). The ultimate (i.e., final) recipient of the analog sound wave is not a computer, but rather the human ear, which we know has limited precision/fidelity. This precision/fidelity exhibits a quantization (or discretization) effect that gives the human auditory system a level of tolerance against noise, such that $\text{message}_1$ + noise = $\text{message}_1$ as long as the magnitude of the noise is less than the quantization step size tolerance. This is why (continuous) AM, regardless of its noisiness, remains useful for transmitting sound waves for ultimate consumption by the human auditory system. If a very high precision (i.e., near-continuous representation, such as 64-bit or 128-bit representation) computer-to-computer communication were to employ (continuous) AM, it would result in catastrophic communication failure, with 100% certainty, in line with the interpretations of the above symbol error probability and channel capacity estimates. In other words, AM radio works only because the human auditory system employs a discrete (or quantized) representation. If it were to employ a truly continuous representation, a noisy message will be confounded (or confused) with another message (i.e., $\text{message}_1$ + noise = $\text{message}_2$), and AM radio music would then sound like noise. The effectiveness of this quantization [30] is another indication that the human auditory system actually employs a discrete (or quantized) representation.

Our mathematical derivation is applicable to any modulation scheme since a distance measure between adjacent modulation symbols ($2d$) exists regardless of the modulation scheme (e.g., frequency modulation, amplitude modulation, phase modulation). Therefore, it is also applicable to any commonly hypothesized information-carrying modulation scheme in the brain, such as spike count (i.e., mean firing rate) coding and interspike interval (ISI) coding [31].

## IV. Application of Discrete Representation Results to Existing Neural Coding Hypotheses

In this section, we apply the discrete results to spiking neurons (i.e., neurons that employ action potentials). Information transmission in spiking neurons takes place in channels (i.e., axons) that are continuous and stochastic. This communications process can be modeled as a repeated refresh (retransmission) of the original message signal via Nodes of Ranvier (NoR) on a myelinated axon that serve as boosting stations, repeatedly restoring the action potential signal that propagates along the axon [32]. Hodgkin and Huxley [33] found that these action potentials operate in a binary mode, consistent with the all-or-none principle proposed by Adrian in 1914 [34]. The key outstanding question is, how is information transmitted (i.e., encoded or carried) by these action potential spikes? At present, there are 4 major neural coding hypotheses (i.e., models), all of which employ continuous representations. We demonstrate the implications of discrete representation on these 4 models. We further illustrate how a discrete ISI neural coding scheme can be used to analyze and interpret electrophysiology experimental data.

## A. Overview of Current Approaches to Neural Coding

The 4 major approaches to modeling neural spikes recorded in the brain (e.g., via electrophysiology experiments) are [31]: mean firing rate as a spike count averaged across time; mean firing rate as a spike density; mean firing rate as a population activity; and, interspike Interval (ISI).

### Mean firing rate as a spike count averaged across time

The most commonly used approach is a spike count averaged across a time window (i.e., temporal average) ([31], section 7.2.1). A time window (or interval) is set and the number of neural spikes produced by a single neuron (in response to an experimental stimulus) within this time window is counted, and then, divided by the duration of the window. For example, in Figure 3, 4 neural spikes were produced by a single neuron within the time window, T; if T was 100 ms, then, the spike count (i.e., mean firing rate) would be 4/(100 ms) = 40 Hz.

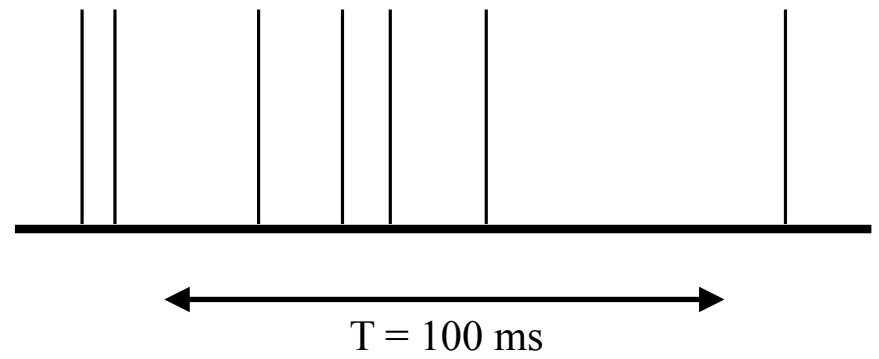

***Figure 3:*** *Mean firing rate as a spike count averaged across a time window (i.e., single neuron, single run).*

Note that, the time window, T, is subjectively set by the experimenter and can vary from one experiment to another, depending on the type of neuron and stimulus. There is no standardized value used across electrophysiology experiments; distinct experiments may use distinct values of T. Typically, T is set to 100 ms or 500 ms, but it could also be set to different (i.e., shorter or longer) durations ([31], section 7.2.1).

### Mean firing rate as a spike count averaged across time and repeated trial runs (i.e., spike density)

The spike density approach employs the same method as the spike count approach, but in addition, the spike count is further averaged over repeated runs (i.e., presentations) of the same stimulus ([31], section 7.2.2). For example, in Figure 4, the same stimulus (i.e., input) is presented to a single neuron repeatedly during 3 different runs. During the first run, the neuron responded with 4 spikes within the time window, T; on the second run, the same neuron responded with only 3 spikes; and on the third run, the same neuron responded with 5 spikes. This run-by-run variability is common in data collected from electrophysiological experiments. If the time window, T, is 100 ms, then, the spike count is (4 + 3 + 5) / (3 * 100 ms) = 40 Hz. Sometimes, instead of using T (i.e., time window), a histogram bin width, Δt, is used in order to plot the neural response as a Peri-Stimulus-Time-Histogram (PSTH) (for details, see [31], section 7.2.2).

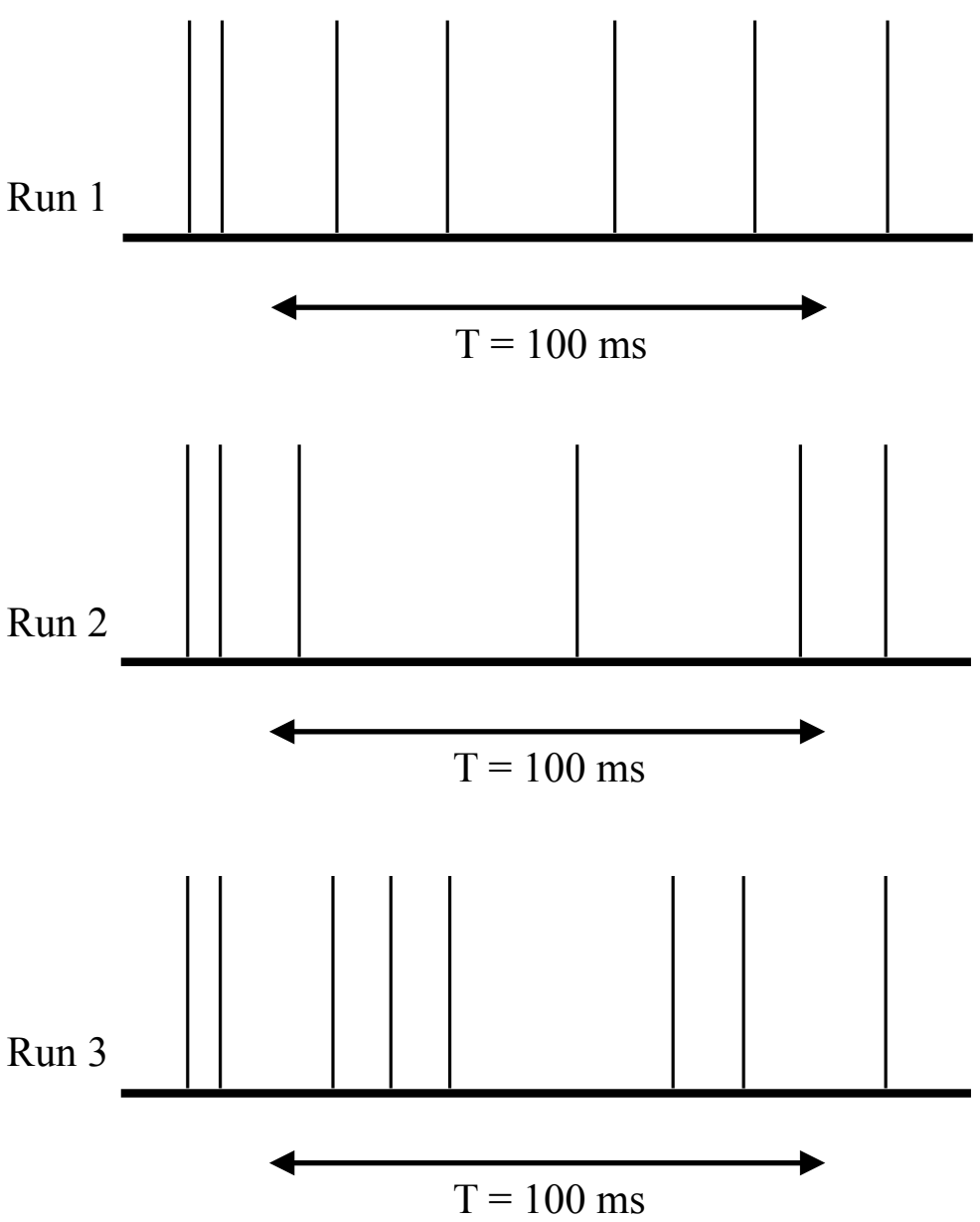

***Figure 4:*** *Mean firing rate as a spike density, averaged across repeated experimental runs (i.e., single neuron, several runs).*

### Mean firing rate as a population activity (i.e., averaged across several neurons)

Under the population activity approach, the spike count is averaged across several neurons that (are believed to) have similar properties (i.e., perform similar neural functions) ([31], section 7.2.3). Here, a single stimulus is presented and the neural spike responses of several functionally equivalent neurons are simultaneously recorded, and then, averaged. For example, in Figure 5, neuron 1 responded with 4 spikes within the time window; neuron 2 responded with 5 spikes; and, neuron 3 responded with 3 spikes. In essence, this approach is similar to the spike density approach, except that multiple neurons with a single stimulus presentation are employed in the averaging process instead of multiple runs of the same stimulus with a single neuron. Again, the neuron-by-neuron variability is common in electrophysiological experiments.

### Interspike interval (ISI)

Under the interspike interval (ISI) approach, it is the temporal delay (i.e., time differential) between 2 successive neural spikes that is used as a neural coding parameter ([31], section 7.3). Typically, a stimulus is presented to a single neuron and the spike train is measured, after which the temporal delay (i.e., time differential) between successive spikes is calculated (see Figure 6) and often analyzed using a histogram. Berger and Levy describe this ISI approach as a *"continuous-time version of differential pulse position modulation"* (DPPM) (see [35], footnote 13).

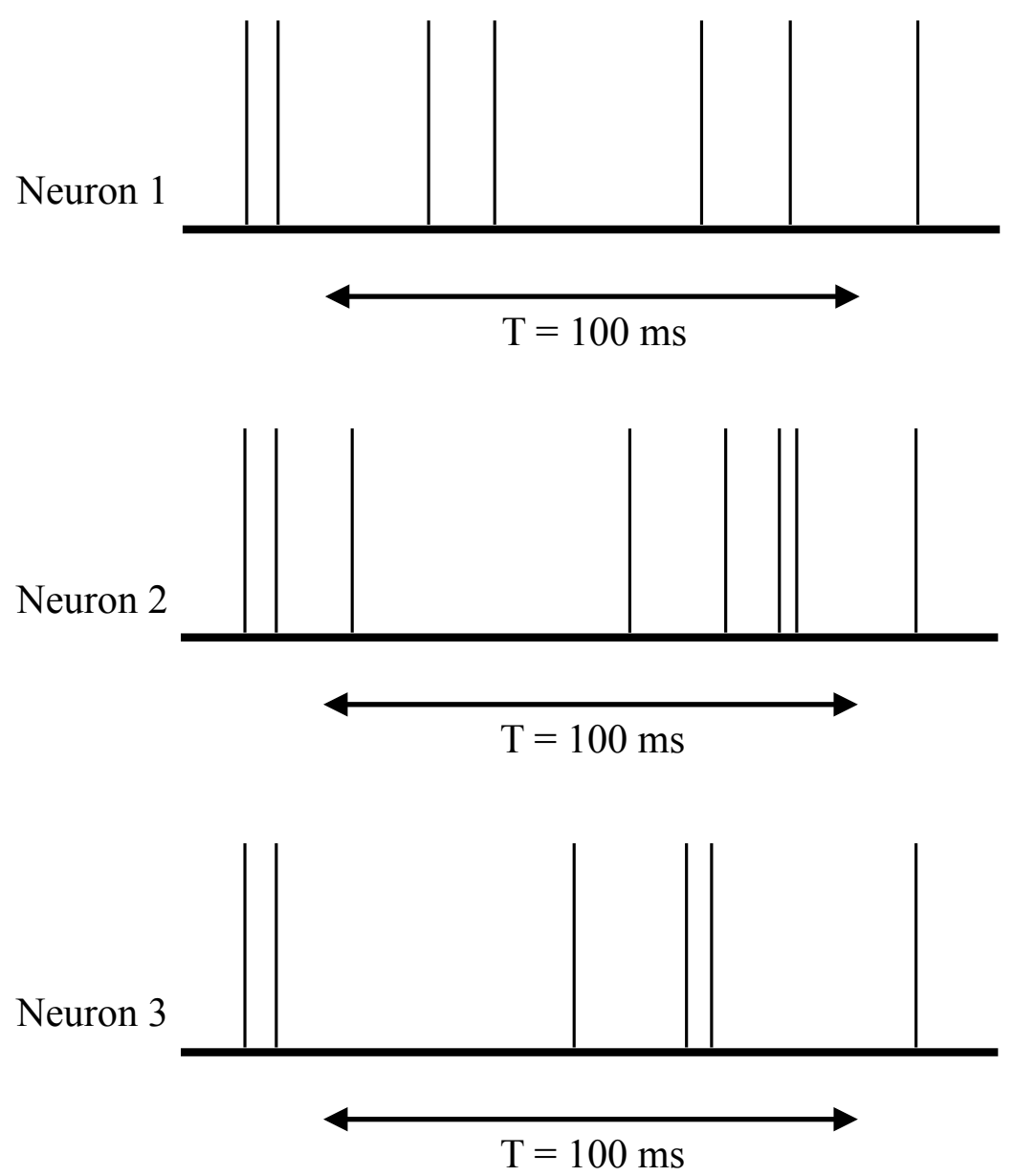

***Figure 5:*** *Mean firing rate as an average across population activity (i.e., multiple similar neurons, single run).*

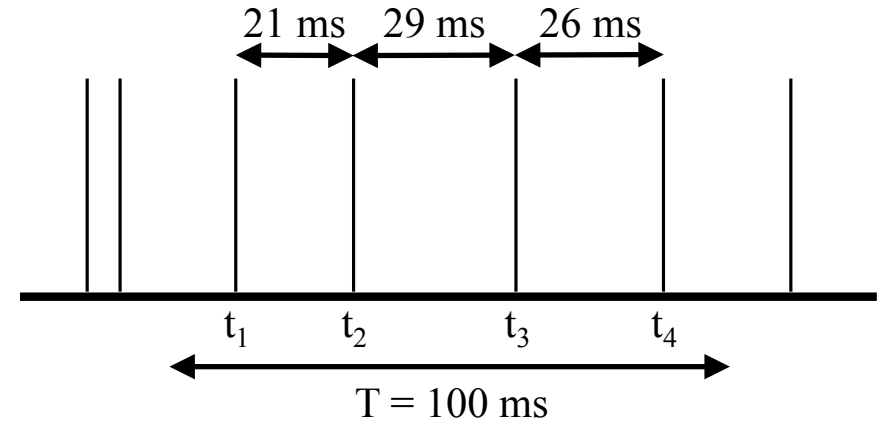

***Figure 6:*** *Interspike interval (ISI) approach to neural coding.*

### *B. Application of Discrete Representation to the ISI Neural Coding Scheme*

All the foregoing neural coding hypotheses assume a continuous representation of information in the brain. In contrast, the work in Section III on symbol error probability and channel capacity estimates shows that continuous representation of information in the brain is not plausible, due to the presence of neural noise [20]. Here, we apply a discrete representation to the 4 neural coding hypotheses and illustrate the implications. We begin by applying discrete representation to the ISI approach. We then extend the approach to the firing rate models.

We approach this by assuming that a discrete representation is employed by default, rather than continuous representation. Assume that we have an M-PAM system as we did in Section III. Suppose that we are using a 8-PAM system (cf. Figure 2), with a signal constellation consisting of the symbol set {-21, -15, -9, -3, 3, 9, 15, 21} (i.e., this modulation scheme has 8 possible transmitted values), communicated over the AWGN channel. If the modulation symbol (or value) -15 was transmitted repeatedly, the received signals will almost never be exactly -15, due to perturbation by AWGN and imperfections in the receiver (e.g., amplifier noise). Therefore, in order to mitigate and/or minimize the effects of AWGN and imperfections, we introduce an error tolerance factor as follows:

$$r = t \pm \Delta$$

where $r$ is the received signal, $t$ is the transmitted signal and $\Delta$ is an error tolerance factor. In our example, $\Delta$, as determined by signal detection theory [36] [37] [38], is equal to half the distance between successive modulation symbols:

$$\Delta = (21 - 15) / 2 = 3$$

This error tolerance approach is equivalent to the discrete M-PAM example in Section III where $2d$ represents the minimum distance between adjacent modulation symbols. Incidentally, this approach is conceptually similar to quantization [39] where the authors used $\Delta$ to represent the spacing between adjacent quantization levels (see Figure 2 of [39]). In fact, quantization was first introduced in 1948 as a form of modulation (Pulse Code Modulation) [30]. As long as the AWGN and imperfections are less than the error tolerance of ± 3 (i.e., margin for error), then, transmission will be error free. For example, if a -15 transmitted signal were perturbed by AWGN and imperfections of -2.98, then, the received signal would be -17.98. By signal detection theory [36] [37] [38], this noisy received signal will be demodulated as a -15 message. However, if the noise and imperfections extend beyond the ± 3 range, say, a perturbation of -3.01, then, the received signal would be -18.01, which would be demodulated as a -21 message, resulting in a communications error.

We can apply this discrete M-PAM example to the brain, by using a discrete ISI representation as the neural coding (i.e., modulation) scheme. Suppose the task is to send a discrete 25 ms ISI neural spike signal (i.e., a 25 ms message) from neuron A to neuron B. What would happen? Well, we know that the brain includes a number of noise sources (e.g., sensory noise, cellular noise, electrical noise, motor noise, synaptic noise) [20]. We also know that the brain is not a flawless neurobiological machine. (see [20], Figure 2b, variability of spike characteristics in repeated trials). This is not too different from the discrete 8-PAM scheme above. With 100% certainty, we know that the transmitted signal (i.e., 25 ms ISI message sent by neuron A) will be perturbed by the noise sources and imperfections in the brain, such that the received signal (i.e., the signal that arrives at neuron B) will very unlikely be a perfect and discrete 25 ms ISI message. (We note that the received signal is a real number, but not all values in this real numbered received signal represent unique messages). Given the random/stochastic nature of neural noise, the received ISI message will vary from one experimental run to another, or from one neuron to another similar neuron. However, as in the discrete 8-PAM case above, as long as the noise and imperfections are less than the error tolerance factor, $\Delta$, reliable error free communications is possible.

As a hypothetical example, suppose that $\Delta = \pm 5$ ms. This means that if the perturbations are less than 5 ms, then, error free communications is possible. Any received signal that falls within the following range of variability would be error free:

$$\begin{aligned} r &= t \pm \Delta \\ &= 25 \text{ ms} \pm 5 \text{ ms} \\ &= (25 \text{ ms} - 5 \text{ ms}, 25 \text{ ms} + 5 \text{ ms}) \end{aligned}$$

= (20 ms, 30 ms)

Reverting to the ISI neural signals depicted in Figure 6, the 3 ISI values are 21 ms, 29 ms and 26 ms. These 3 values all fall within the range of (20 ms, 30 ms), such that these 3 ISI signals would be demodulated as 25 ms messages if 25 ms was indeed the actual transmitted message and the error tolerance, Δ, was ± 5 ms. (See Figure 7). If, however, the Δ was only ± 3 ms, then, the tolerance range reduces to (22 ms, 28 ms) such that only the third ISI value (i.e., 26 ms) would be correctly demodulated as a 25 ms transmitted signal. Therefore, error tolerance is a way of mitigating and/or minimizing stochastic variability and inconsistencies that arise from neural noise sources and/or neurobiological imperfections.

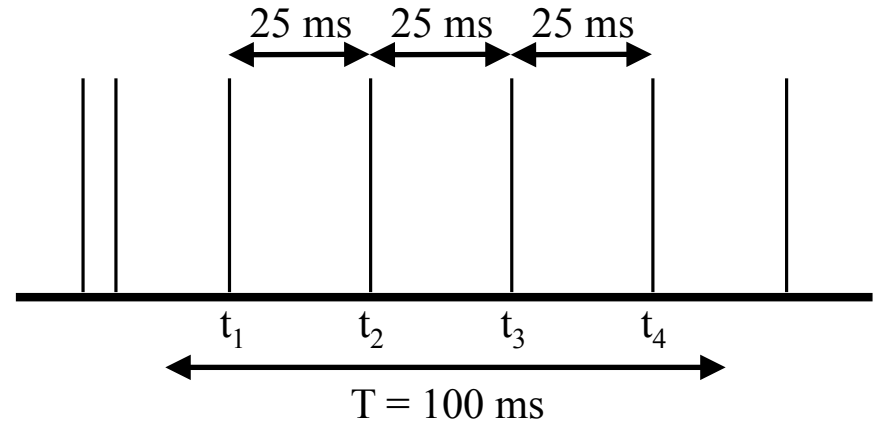

***Figure 7:*** *Likely intended transmitted discrete ISI signals given the real-valued received ISI signals as depicted in Figure 6, assuming an error tolerance, Δ = ± 5 ms.*

For completeness, we now suppose that a continuous representation is employed in the brain. In such a case, every possible value of the ISI signal represents a distinct message. For example, an ISI signal of 25.00001 ms would carry a different message than an ISI signal of 25.00000 ms. Therefore, in the case of continuous representation, the error tolerance factor, Δ, is actually zero. Consequently, at the destination neuron, there is no room (or margin) for error that might arise from neural noise and/or neurobiological imperfections; all received signals, except for an exact 25 ms signal, would be demodulated incorrectly. It is, therefore, neither possible nor plausible for information in the brain to be represented in continuous form; it has to be represented in discrete form.

One major implication of this discrete conclusion is that the likely modulation system employed by the brain is discrete-time (i.e., "ordinary", digital) differential pulse position modulation (DPPM) [40], as opposed to a continuous time-version of DPPM as hypothesized by Berger and Levy [35].

### *C. Application of Discrete ISI Neural Coding to Analyzing Electrophysiology Experimental Data*

In this subsection, we illustrate how a discrete ISI neural coding scheme can be applied to the analysis and interpretation of neural spike data. Here, we use the electrophysiology experimental data of one animal subject (i.e., a ferret) from a wider study [41].

Recently, it was discovered that individual neurons (i.e., cerebellar Purkinje cell neurons) can be trained to respond to an artificial conditioned stimulus (CS) [42] [43]. Specifically, individual neurons are shown to be able to remember the time interval (i.e., duration) of a stimulus. Prior to training (i.e., under normal circumstances), the neuron does not respond any differently to an artificial stimulus; it continues to fire at its usual patterns. However, after training, the neuron learned to respond by pausing its neural firings for a duration that is proportionate to the duration of the artificial stimulus. This neural firing pause has approximately the same duration as that of the artificial stimulus.

Figure 8 shows the raster plot of a neuron that has been trained to respond to an artificial timing stimulus [41]. Each black dot represents a neural spike. The 2 blue vertical lines mark the onset and the end of the artificial timing stimulus. The stimulus is 200 ms in duration. The y-axis is the trial number, whereby each horizontal line of black dots represents the neural spike firing patterns of the trial in question. Twenty post-training trials are plotted in Figure 8. The x-axis represents the time stamp, where time 0 is marks the onset of the stimulus. What is observed is the spike pausing effect [42]: in each trial, the neuron pauses its firing when the stimulus is present, and the duration of this pause is approximately the same as the duration of the stimulus. What is also clearly observable is the variability of the pause duration from one trial to another.

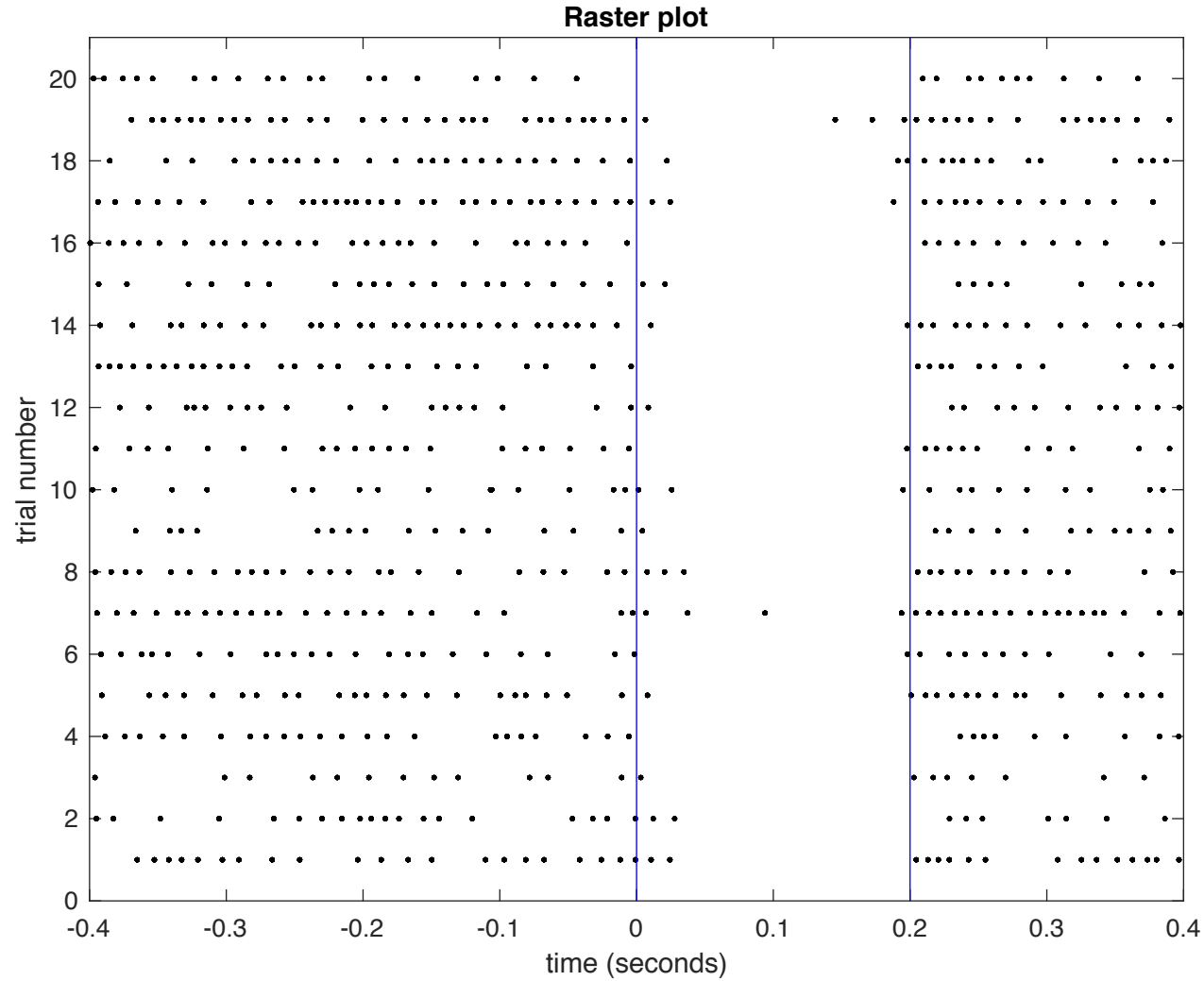

***Figure 8:*** *Raster plot of a neuron that has been trained to respond to an artificial timing stimulus of 200 ms in duration [41].*

One plausible interpretation of this raster plot is that the neuron intended to pause for 200 ms during each trial, in response to the 200 ms stimulus. However, given the presence of neural noise and/or neurobiological imperfections, the neuron is unable to produce an exact 200 ms (i.e., intended pause) every single time, thereby, resulting in the trial-by-trial variability (i.e., actual pause). These pauses fall within a range of values. In a manner equivalent to a discrete ISI neural coding scheme in Section IV-B, we can model this variability using an error tolerance factor:

$$\textit{actual pause} = \textit{intended pause} \pm \Delta$$

The pause duration for each trial can be obtained by calculating the time differential (i.e., ISI) between the 2 spikes (that created the pause). Since there are 20 trials, there are 20 pause ISIs. Figure 9 shows the histogram and the empirical cumulative distribution function (CDF) of these pause ISIs. The black vertical dashed line on the histogram shows the duration of the stimulus (i.e., 200 ms). The mode is 200 ms, with a mean of 192.5 ms. The median (i.e., $50^{th}$ percentile), as shown by the green dashed line (on the empirical CDF) is 199.8 ms.

In order to illustrate how we can use discrete ISI neural coding to analyze this data, we apply an error tolerance factor of 25 ms. This gives us:

$$\textit{actual pause} = 200 \text{ ms} \pm 25 \text{ ms} = (175 \text{ ms}, 225\text{ms})$$

This range is shown by the 2 blue dotted lines in the empirical CDF, where the lower boundary (i.e., 175 ms) represents the $30^{th}$ percentile and the upper boundary (i.e., 225 ms) represents the $90^{th}$ percentile. This covers 60% of the data. Clearly, if a wider error tolerance was employed (e.g., 50 ms), a larger proportion of the data will be covered. What we are attempting to illustrate here is that, by using a discrete ISI neural coding scheme with an error tolerance of 25 ms, the trial-by-trial variability of the neural spike pauses (in response to the 200 ms stimulus) can be easily modeled and explained from a communications systems engineering perspective.

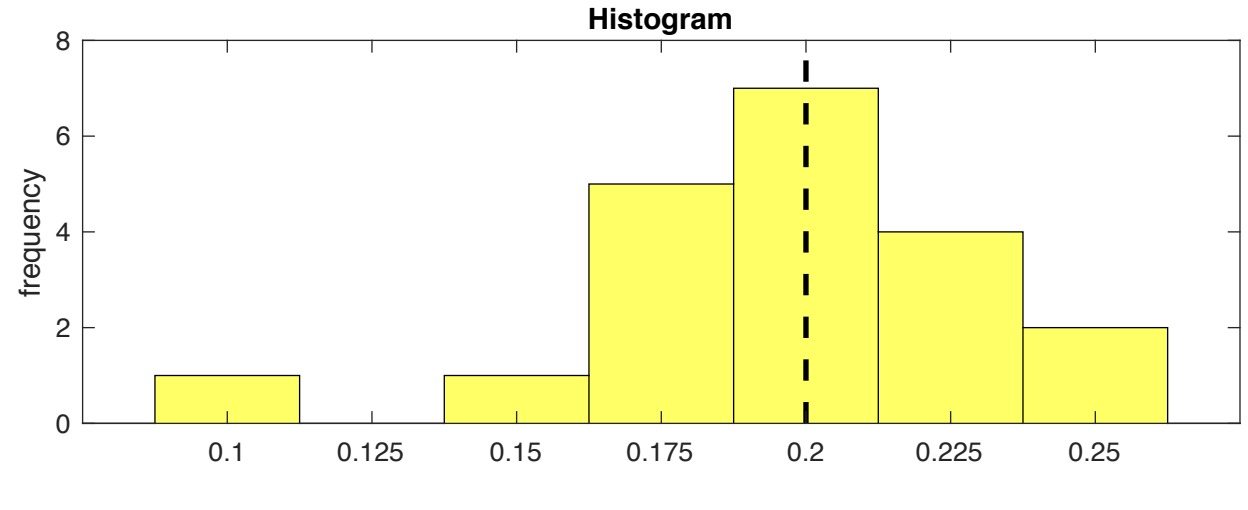

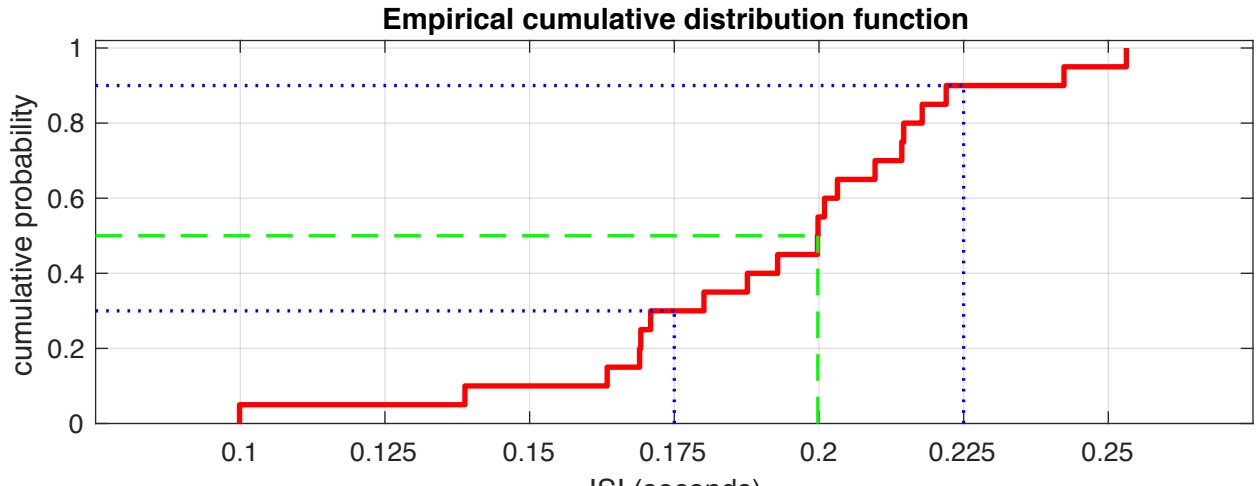

***Figure 9:*** *Histogram and empirical cumulative distribution function of the 20 pause ISIs.*

*D. Application of Discrete Representation to the 3 Mean Firing Rate Coding Schemes*

We begin by examining the case where the brain employs continuous representation of information. In such a situation, every possible value of the firing rate represents a distinct message. For example, a firing rate of 40.00001 Hz would carry a different message than a firing rate of 40.00000 Hz. Therefore, similar to the case of ISI neural coding, the error tolerance factor, $\Delta$, is actually zero if a continuous representation is employed, resulting in no room (or margin) for error that might arise from neural noise and/or neurobiological imperfections. It is, therefore, neither possible nor plausible for information in the brain to be represented in continuous form.

One could argue that a neural firing rate code may not employ such a large number of decimal places (i.e., 5 decimal places) in terms of precision – that, perhaps it operates using only 1 decimal place of precision. If that were indeed the case, then, 40.11 Hz would carry the same message as 40.10 Hz since both are the same signals if the precision is truncated to only 1 decimal place (i.e., 40.1 Hz). However, such a 1-decimal-place precision is no longer a continuous representation; the representation is not a real number but rather a discretized or quantized form. Again, there is no case for the brain to employ continuous representation in a firing rate neural coding scheme.

All of the firing rate models (i.e., spike count, spike density and population activity) employ an averaging process. One consequence of averaging is that the run-by-run or neuron-by-neuron variability is no longer a problem. This approach "takes care" of the variability by averaging things out. As a result, the models' averaged outputs are essentially indifferent and equivalent to equally spaced neural spikes. For example, the resultant post-averaged spike trains of Figures 9, 10 and 11 are essentially reduced (by the averaging process) to the same uniform spike train whereby neural spikes are spaced equally, as shown in Figure 10.

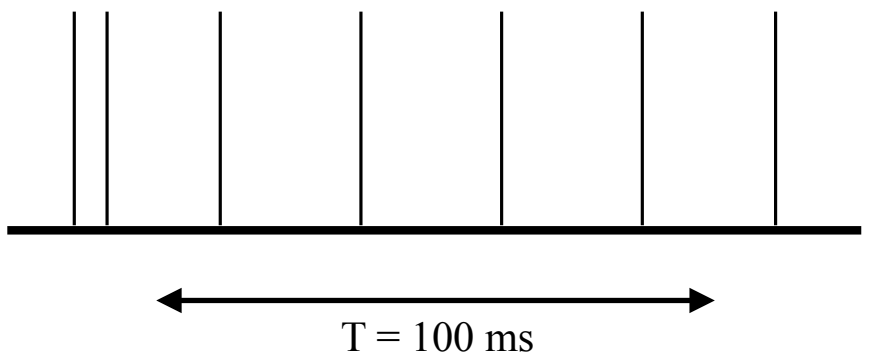

***Figure 10:*** *Example of a uniform spike train whereby neural spikes are spaced equally.*

Since only the mean of the 3 firing rate models is useful (as opposed to the exact characteristic of each spike train), this offers another hint that continuous representation is not actually employed. That is, if continuous representation was employed, then, every unique value (or signal) carries a distinct message (information) and the averaging process would confound messages. However, since only the average is useful, then, this implies that certain ranges of the signal values are indifferent to (i.e., carry the same message as) the message carried by the mean value. Specifically, spike count values located just below and just above the mean would be averaged out, reverting the outcome towards the mean. For example, in Figure 4, run 1 has 4 spikes, run 2 has 3 spikes and run 3 has 5 spikes, resulting in an average of 4 spikes (within the 100 ms time window). Here, $\Delta = \pm$ 1 spike, such that 4 spikes $\pm$ 1 spike covers the range [3, 5] spikes, allowing room for error. It is plausible that any spike count that falls inside such a range conveys the same message. This example, though hypothetical and conveniently illustrated, fits well with the use of an error tolerance, $\Delta$, (i.e., discretization, quantization, digital modulation schemes) in order to mitigate and/or minimize neural noise and/or neurobiological imperfections. If continuous representation was employed, then, noise and/or imperfections would have a highly destructive effect.

### *E. Equivalence of the 3 Mean Firing Rate and ISI Neural Coding Schemes*

The conventional ISI neural coding approach (i.e., continuous representation) is often analyzed using a histogram ([31], section 7.3). The construction of a histogram involves a binning process, which divides the entire range of data values into equally sized intervals (or buckets). This binning process is, in essence, a quantization process. Consequently, any analysis that involves a histogram is essentially and inadvertently quantizing (binning) the continuous data, turning the data and the results into a discrete form. Histograms are also commonly used in conjunction with one of the 3 mean firing rate models (i.e., spike density) by way of a Peri-Stimulus-Time-Histogram (for details, please see [31], section 7.2.2). Furthermore, the averaging process of the 3 mean firing rate neural coding schemes results in the same (or indifferent) uniform spike train where neural spikes are spaced equally, as shown in Figure 10. Such a uniformly spaced spike train (due to the averaging process) is conceptually very similar to the discrete ISI spike train of Figure 7 (i.e., quantized ISI). Our sense is that, if a discrete representation was employed, then, all 4 schemes are essentially one and the same. In our prior examples, the discretized (i.e., quantized) mean firing rate models become:

$$f \pm \Delta_f = Q_f[f]$$

where f is the firing rate, $\Delta_f$ is the error tolerance of the firing rate and $Q_f[.]$ denotes a quantization (or discretization) process with a uniform spacing of $2\Delta_f$. In the case of discrete ISI neural coding:

$$ISI \pm \Delta_I = Q_I[ISI]$$

where $\Delta_I$ is the error tolerance of the ISI and $Q_I[.]$ denotes a quantization (or discretization) process with a uniform spacing of $2\Delta_I$. Since frequency is the inverse of time, then:

$$Q_f[f] = 1/Q_I[ISI]$$

In this way, a discrete firing rate of 40 Hz is equivalent to the inverse of a discrete 25 ms ISI:

$$Q_f[f] = 40 \text{ Hz} = 1/(25 \text{ ms}) = 1/Q_I[ISI]$$

Discretization (or quantization), by way of introducing an error tolerance factor, mitigates and/or minimizes neural noise and/or neurobiological imperfections, while at the same time, it reconciles and unifies the current 4 major approaches (i.e., hypotheses) to modeling neural spike signals.

## V. Extension of Discrete Representation Results to Weber's Law

In 1834, Ernst Weber discovered that *"the sensitivity of a sensory system to differences in intensity depends on the absolute strength of the stimuli"* [44]. Specifically, the sensitivity to differences/changes to a stimulus is described by Weber's Law as:

$$\frac{\Delta I}{I} = k$$

where $I$ is a reference stimulus, $\Delta I$ is the minimum amount that the stimulus intensity must be changed (e.g., increased) in order for the difference (e.g., increment) to be detectable (i.e., discriminated) by the (human) sensory system (also known as Just Noticeable Difference (JND) or Difference Threshold), and $k$ is a constant (also known as Weber's fraction or Weber's constant). In other words, the size of the Just Noticeable Difference (JND) (i.e., $\Delta I$) is a constant proportion (i.e., Weber's fraction) of the reference stimulus:

$$\Delta I = kI$$

For example, the Weber fraction for lifted weight (i.e., heaviness) is 0.020 [45]. This means that, in order for the weight change to be detectable by the human sensory system, the change must be at least 0.020 of the reference weight (i.e., 2% of the reference weight). Suppose the reference weight is 10 kg. Weber's Law states that any change (i.e., increment) that is less than 200 g (i.e., 2% of 10 kg) cannot be detectable if it was added to the reference 10 kg weight. In order for a change to be detectable, the increment must be greater than or equal to 200 g. The significance of Weber's Law lies in its scalability. For example, suppose the reference weight is now 100 g. According to Weber's Law, change is only detectable if the increment is greater than or equal to 2 g (i.e., 2% of 100 g). This means that a human cannot tell the difference between 101 g and 100 g; a difference can only be detected between 102 g and 100 g. A curious reader may wonder whether or not 1 g is actually detectable at all by the human sensory system regardless of the reference weight. The answer to this question is yes (1 g is the weight of a 5 carat diamond). Weber's Law is applicable to a wide range of sensory stimuli, such as brightness, loudness, taste (of salt), electric shock and vibration [45], each with its own distinct Weber's fraction.

Our novel insight here is that the JND is approximately equivalent to the error tolerance factor, Δ, that we introduced in Section IV. A change of at least $+\Delta I$ is detectable, and likewise, so is a change of at least $-\Delta I$. In contrast, a change of less than $+\Delta I$ or less than $-\Delta I$ is not detectable. In other words, changes that fall within the $(I - \Delta I, I + \Delta I)$ range are undetectable by the human sensory system (i.e., No Noticeable Difference) and therefore, are detected indifferently to $I$. That is, any value that falls within $I \pm \Delta I$ is detected as $I$. The remarkable resemblance between Weber's Law and the error tolerance factor applied to the neural coding of Section IV (i.e., quantization, discretization, digital modulation scheme) is uncanny, and unlikely to be a coincidence. Could the range of undetectable differences in Weber's Law have arisen due to a corresponding range of error tolerance in neural coding? We posit that a direct link between discrete neural coding (i.e., ISI ± Δ) and Weber's Law (i.e., $I \pm \Delta I$) is both possible and plausible. Like discrete neural coding, our sense is that Weber's Law is a manifestation and consequence of the discrete representation of information in the brain, possibly by way of non-linear quantization, similar to the results on quantized representation of probability in the brain [46].

## VI. Conclusions

We have performed a detailed study of continuous-versus-discrete information representation in the brain using basic tenets of communications systems engineering. Our symbol error probability and channel capacity analyses show that a continuous modulation scheme is neither plausible nor feasible. We have also applied the discrete representation results to the 4 common neural coding schemes, ruling out the plausibility of continuous representation. In addition, we have demonstrated how a discrete ISI neural coding scheme can be applied to the analysis and interpretation of electrophysiology experimental data. Our extension of the discrete representation to Weber's Law highlights the similarity between JND and quantization, offering a potential way to directly connect discrete neural coding to Weber's Law. Overall, our discrete conclusion in this paper signifies a game changer to the field of neuroscience.

Going forward, we believe that the correct research question is no longer that of continuous-versus-discrete, but rather, how fine grained the discreteness is (i.e., how many bits of precision). It is very plausible that different parts of the brain (e.g., visual, auditory, cognitive decision-making) operate at different levels of discreteness, possibly based on different numbers of quantization levels. For example, camera photographs are commonly encoded in 24-bit RGB color (i.e., 16,777,216 distinct colors) whereas CD music is commonly encoded in 16-bit audio (i.e., 65,536 distinct levels of loudness), representing (or approaching) the limits of the brain's visual and auditory precision/discreteness capabilities respectively.

In terms of neural coding, our view is that a discrete ISI neural code is essentially equivalent to a discrete firing rate neural code, with one being the inverse of the other. Certain ranges of ISI (or frequency) values are very likely to represent the same information message. One consequential question is regarding the size of the error tolerance, $\Delta$. That is, how widely spaced are two adjacent modulation symbols? Another curious question is the size of the signal constellation. The English language has an alphabet size of 26 symbols (i.e., A to Z). If discrete-time differential pulse position modulation (DPPM) [40] is indeed the modulation scheme employed, how many distinct modulation symbols are there in the brain's neural "language" (i.e., signal set)? As an extension, another pertinent question is whether adjacent symbols are equally/uniformly spaced or spaced in a non-linear manner. That is, as ISI values (or symbols) become larger (e.g., from 50 ms to 200 ms), $\Delta$ may also become larger (e.g., from 5 ms to 20 ms), possibly in a similar manner to Weber's Law. It is known that Weber's Law is optimal for many sensory systems [47]. Moreover, the Weber-Fechner Law further states that there is a logarithmic relationship between perceived intensity and actual stimulus [48], and this non-linear relationship has been found to be optimal for all stimulus distributions [14].

We end this paper with a possible starting point in the search for answers. The most conventional frequency bands used to identify functional brain networks are Delta (1–4 Hz), Theta (4–8 Hz), Alpha (8–13 Hz), Beta (13–30 Hz), Gamma (30–80 Hz), and High Gamma (80–150 Hz) [49]. These frequency bands are plotted in Figure 11, both in a linear scale and a logarithmic scale. Each frequency band could potentially represent a crudely sized category (or group) of (modulation) symbols, within which there may exist more fine spaced (modulation) symbols. Such (modulation) symbols can be probed and uncovered using a careful and systematic selection of experimental stimuli.

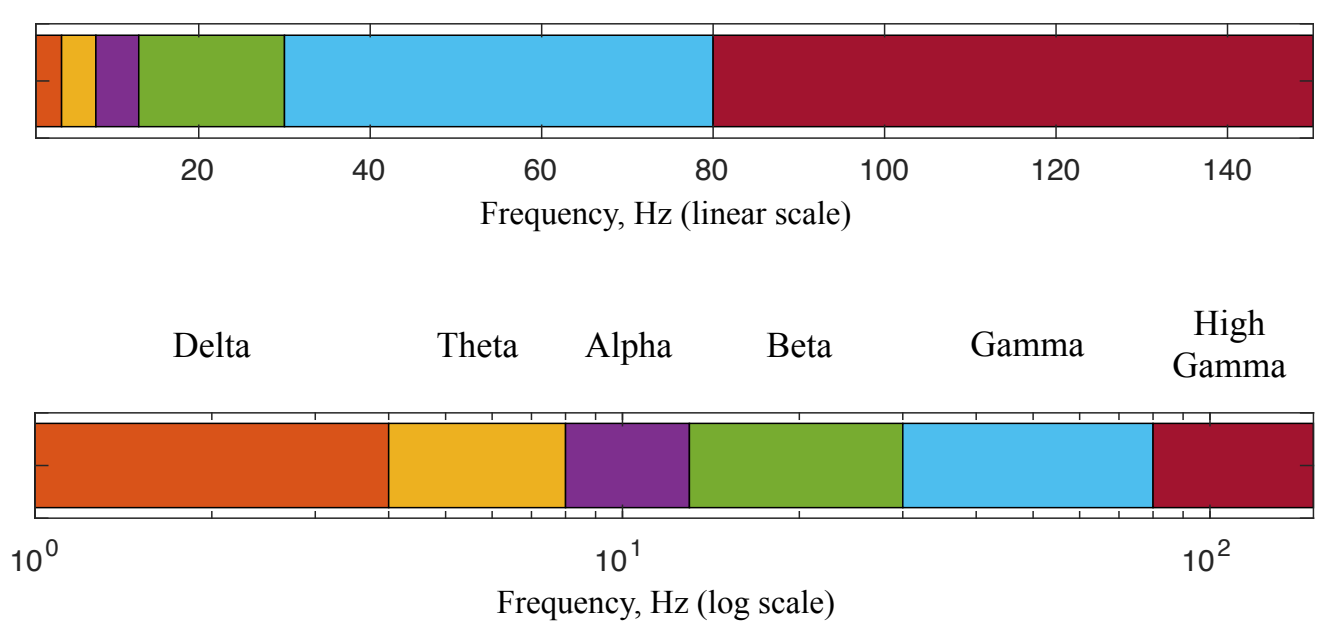

***Figure 11:*** *The most conventional frequency bands used to identify functional brain networks [49], plotted on a linear scale (top graph) and a logarithmic scale (bottom graph).*

## Acknowledgment

J. Tee thanks C. R. Gallistel and M. Woodford for their guidance and advice. J. Tee and D. P. Taylor thank the late W. H. Tranter for his constructive comments and suggestions. J. Tee thanks F. Johansson of the Hesslow Lab at the Department of Experimental Medical Science at Lund University (Sweden) for sharing his experimental data.

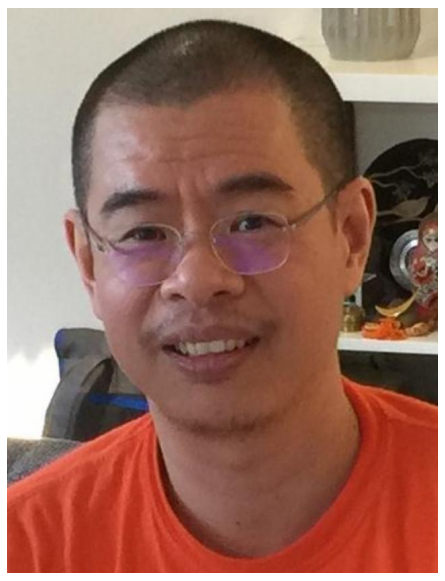

James Tee (M'17) completed his Ph.D. in Electrical & Electronic Engineering at the University of Canterbury in 2001, where he worked on Turbo Codes under the supervision of Des Taylor. Subsequently, he held various industry and policy positions at Vodafone Group, the World Economic Forum, New Zealand's Ministry of Agriculture & Forestry, and the United Nations. To facilitate his career transitions, he pursued numerous supplementary trainings, including an MBA at the Henley Business School, and an MPhil in Economics (Environmental) at the University of Waikato. In 2012, James began his transition into scientific research at New York University (NYU), during which he completed an MA in Psychology (Cognition & Perception) and a PhD in Experimental Psychology (Neuroeconomics). Afterwards, he worked as an Adjunct Assistant Professor at NYU's Department of Psychology, and a Research Scientist (Cognitive Neuroscience) at Quantized Mind LLC. Since 2017, he is an Adjunct Research Fellow at the University of Canterbury. James is an Eastern medicine physician, with an MS in Acupuncture from Pacific College of Oriental Medicine. In August 2020, he began further training in Substance Abuse Counseling at the New School for Social Research (NSSR). His current research interests in neuroscience focuses on reverse engineering the communications codebook (i.e., signal constellation) of the Purkinje cell neuron. James is also working on Artificial Intelligence (AI) approaches inspired by insights drawn from psychology (cognition, perception, decision-making) and neuroscience.

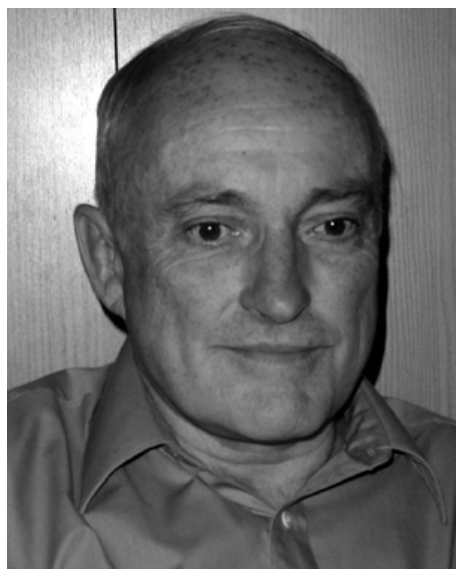

Desmond P. Taylor (LF'06) received the Ph.D. degree in electrical engineering from McMaster University, Hamilton, ON, Canada, in 1972. From 1972 to 1992, he was with the Communications Research Laboratory and the Department of Electrical Engineering, McMaster University. In 1992, he joined the University of Canterbury, Christchurch, New Zealand, as the Tait Professor of communications. He has authored approximately 250 published papers and holds several patents in spread spectrum and ultra-wideband radio systems. His research is centered on digital wireless communications systems focused on robust, bandwidth-efficient modulation and coding techniques, and the development of iterative algorithms for joint equalization and decoding on fading, and dispersive channels. Secondary interests include problems in synchronization, multiple access, and networking. He is a Fellow of the Royal Society of New Zealand, the Engineering Institute of Canada, and the Institute of Professional Engineers of New Zealand.